# A note on the estimation of the mass of the universe

P. R. Silva – Departamento de Física – ICEx – Universidade Federal de Minas Gerais
C. P. 702 – 30123-970 – Belo Horizonte – MG – Brazil
e-mail : prsilvafis@terra.com.br

## ABSTRACT

We make an estimation of the mass of the universe by considering the behavior of a very special test particle when described both by using the Newtonian mechanics as well through a scalar field theory of the Yukawa kind. Naturally, Hubble's law is also taken in account.

The determination of the critical density of matter that halts the expansion of the universe can be done within the paradigm of the newtonian mechanics, by using the conservation of the mechanical energy of a test particle and neglecting the contribution of the cosmological constant term [1, 2]. In these simple calculations we consider that the test particle suffers the gravitational influence of the amount of matter contained inside a sphere of radius r. This attractive character is balanced by a repulsive-like term which accounts for the expansion of the universe as stated by the Hubble's law [3, 4]. The center of this sphere is sited anywhere in the universe.

However let us consider a "very special" test particle of mass m. We are taking this particle as probing all the mass M of the observed universe, represented by a sphere of radius R, with its mass homogeneously distributed through it. Therefore we can write the energy of the interaction of the test particle with the universe as

$$U = - (GMm)/R, \qquad (1)$$

where G is the gravitational constant.

The very special character of this test particle will be attributed to it by making the requirement

$$GMm = GM_{Pl}^2 = \alpha_g \hbar c = \hbar c. \qquad (2)$$

In (2), $M_{Pl}$ is the Planck mass and $\alpha_g$ is the "gravitational running coupling constant", which is taken to be equal to the unity at the energy scale of the Planck mass.

It is convenient to rewrite equation (2) as

$$Mm = M_{Pl}^2 = (\hbar c)/G. \qquad (3)$$

While the gravitational coupling constant is an increasing function of the mass (energy), it is well known that the running coupling constant of the strong force $\alpha_s$ exhibits the property of the asymptotic freedom, being a decreasing function of the energy [5, 6, 7]. At the energy scales of the nucleon mass, the strong coupling $\alpha_s$ approaches to the unity.

Quantum gravity demands that to the particle which intermediates the gravitational coupling, the graviton, must be attributed a spin of 2 [8, 9]. On the other hand, the most fundamental description of the strong interaction, the quantum chromodynamics (QCD) [10, 11], requires that quarks: particles endowed with color charges (red, green, blue), interact through the exchange of gluons, quanta of a field having spin 1.

However, before the advent of QCD a tentative theory for the strong interaction was proposed by Yukawa, where nucleons (proton and neutron) interact by the exchange of massive particles of spin zero called pion.

In this note we propose that a Klein-Gordon-Yukawa equation could be used to describe gravity at large distances. Then we write [12]

$$\Delta\Psi - (1/c^2)\partial^2\Psi/\partial t^2 = (mc/\hbar)^2\Psi, \qquad (4)$$

where $\Psi$ is a scalar field associated to the exchange of particles of mass m and spin zero.

The static solution of (4) is

$$\Psi = (g/r)\exp(-r/R), \qquad (5)$$

where

$$R = \hbar/mc, \qquad (6)$$

is the range of the Yukawa field and g is the "strong charge".

It is interesting to observe that if we insert (6) in the potential energy given by (1), we obtain the total energy E, namely

$$E = U(R = \hbar/mc) + mc^2 = 0. \qquad (7)$$

This universe created from nothing was discussed by Tryon [13] some time ago.

Now we turn to Hubble's law [3, 4] that states that for two galaxies separated by a distance r, the velocity of expansion of a homogeneous and isotropic universe [14] is

$$v = H_0 \, r, \qquad (8)$$

where $H_0$ is the constant of Hubble. We are going to use a similar relation as a means to set up a scale for the universe, and write

$$c = H_0 \, R, \qquad (9)$$

where c is the velocity of light in vacuum. Identifying R the "Hubble's radius" of (9) with the range of the Yukawa field given by (6), and inserting this information into relation (3) we get

$$M = c^3/(GH_0). \qquad (10)$$

The estimation for the mass of the universe given by (10) is identical as that proposed before by Sharma and Sharma [15], by using a different reasoning.

The comparison of (6) and (9) also relates the mass energy of this "very special test particle" to the "Hubble's frequency" through the equation

$$mc^2 = \hbar H_0. \qquad (11)$$

This last result has also been obtained by Sharma and Sharma [15] as the lower bound on a particle mass (please see also [16]).

ACKNOWLEDGEMENTS – We are grateful to Domingos Sávio de Lima Soares and Nilton Penha Silva for helpful comments of a previous version of this work.